\documentclass[twocolumn,prl,amsmath,amssymb,showpacs,superscriptaddress]{revtex4-1}
\usepackage{epsf}      
\usepackage{graphicx}
\usepackage{color}
\usepackage{gensymb}
\usepackage{sidecap}

\begin{document}
\title {Magnetocaloric properties and critical behavior of Co$_2$Cr$_{1-x}$Mn$_x$Al Heusler alloys}

\author{Priyanka Nehla}
\affiliation{Department of Physics, Indian Institute of Technology Delhi, Hauz Khas, New Delhi-110016, India}
\author{V. K. Anand}
\affiliation{\mbox{Helmholtz-Zentrum Berlin f$\ddot{u}$r Materialien und Energie GmbH, Hahn-Meitner Platz 1, D-14109 Berlin, Germany}}
\affiliation{Department of Physics, SRM University, AP - Amaravati, 522502, Andhra Pradesh, India}
\author{Bastian Klemke}
\affiliation{\mbox{Helmholtz-Zentrum Berlin f$\ddot{u}$r Materialien und Energie GmbH, Hahn-Meitner Platz 1, D-14109 Berlin, Germany}}
\author{Bella Lake}
\affiliation{\mbox{Helmholtz-Zentrum Berlin f$\ddot{u}$r Materialien und Energie GmbH, Hahn-Meitner Platz 1, D-14109 Berlin, Germany}}
\author{R. S. Dhaka}
\email{rsdhaka@physics.iitd.ac.in}
\affiliation{Department of Physics, Indian Institute of Technology Delhi, Hauz Khas, New Delhi-110016, India}

\date{\today}

\begin{abstract}
We study the magnetocaloric effect and critical behavior of Co$_2$Cr$_{1-x}$Mn$_x$Al ($x=$ 0.25, 0.5, 0.75) Heusler alloys across the ferromagnetic (FM) transition (T$_{\rm C}$). The Rietveld refinement of x-ray diffraction patterns exhibit single phase cubic structure for all the samples. The temperature dependent magnetic susceptibility $\chi$(T) data show a systematic enhancement in the Curie temperature and effective magnetic moment with Mn concentration, which is consistent with the Slater-Pauling behavior. The M(H) isotherms also exhibit the FM ordering and the analysis of $\chi$(T) data indicates the nature of the phase transition to be a second order, which is further supported by scaling of the entropy curves and Arrott plot. Interestingly, the Mn substitution causes an increase in the magnetic entropy change and hence large relative cooling power for multi-stage magnetic refrigerator applications. In order to understand the nature of the magnetic phase transition we examine the critical exponents $\beta$, $\gamma$, $\delta$ for the $x=$ 0.75 sample by the modified Arrott plot and the critical isotherm analysis, which is further confirmed by Kouvel-Fisher method and Widom scaling relation, respectively. The estimated values of $\beta=$ 0.507, $\gamma=$ 1.056, $\delta=$ 3.084 are found to be close to the mean field theoretical values. The renormalized isotherms (m vs h) corresponding to these exponent values collapse into two branches, above and below T$_{\rm C}$ that validates our analysis. Our results suggest for the existence of long-range FM interactions, which decays slower than power law as $J(r)\sim r^{-4.5}$ for a 3 dimensional mean field theory. 

\end{abstract}
 
\maketitle

\section{\noindent ~Introduction}

Heusler alloys have attracted great attention to the scientific community after the discovery of ferromagnetism in Cu$_2$MnSn alloy in 1903 \cite{Heusler1903}. The full Heusler alloys, having $X_{\rm 2}YZ$ formula, crystallize in cubic structure with $Fm\bar{3}m$ space group, where $X$, $Y$ are the transition elements, and $Z$ is the main group element \cite{GrafPSSC11, GalanakisPRB02, AlijaniPRB11}. In recent years, many unique physical properties have been found in Heusler alloys such as shape memory effect \cite{KainumaNature06, DeviPRB18, VasilevPRB99, DhakaSS09}, magnetoresistance (MR) \cite{WenPRA14}, half metallicity, spin filters, and spin injection devices \cite{Wolf01, GrootPRL83, GalanakisJPCM07}. The half metallic ferromagnets belong to a new class of materials, in which the band structure shows semiconducting nature for one spin channel and metallic for other spin channel, and it was first observed in NiMnSb Heusler alloys by Groot {\it et al.} \cite{GrootPRL83}. The observed high magnetic moment, the Curie temperature (T$_{\rm C}$) well above the room temperature and half metallic ferromagnetic (HMF) nature in Heusler alloys (in particular Co-based) make them attractive candidate for spintronics applications \cite{GrafIEEE11, WurmehlAPL06, NehlaJALCOM19, KublerPRB07, GalanakisJPCM07}. In this direction, the ferromagnetic Co$_2$CrAl is particularly interesting because of its many useful physical properties; for example T$_{\rm C}$ (330~K) just above room temperature, theoretically predicted 100\% spin polarization, anomalous Hall conductivity, and the calculated magnetic moment of 3$\mu_{\rm B}$ \cite{HusmannPRB06, KourovPSS13,ElmersPRB03}. Intriguingly, the Co$_2$MnAl Heusler alloy has been also predicted to show half metallic ferromagnet with the T$_{\rm C}$ value much higher (about 665~K) and the total magnetic moment equal to 4$\mu_B$/f.u., which follow the Slater--Pauling behavior \cite{OzdoganPRB06, ChenAPL04, UmetsuJAP08}. The appearance of Weyl points near the Fermi energy may cause the anomalous Hall conductivity in these materials \cite{KublerEPL16}. Therefore, it is vital to study the effect of Mn substitution at Cr site in Co$_2$CrAl to understand the magnetic properties, which play crucial role in practical applications \cite{GalanakisJPCM04, KudrnovskyPRB13}. 

Apart from the HMF nature, magnetocaloric effect (MCE) is one of the most interesting properties found in Heusler alloys, which is useful in magnetic refrigerators \cite{ZhangSR18}. The MCE is a phenomena where a material heats up or cools down on the application or removal of the magnetic field. Initially, rare earth materials were considered to be the best candidates for MCE because of their high value of magnetic entropy change ($\Delta S_m$). For example, ferromagnetic element Gd which has a very high magnetic moment shows T$_{\rm C}$ around room temperature and -$\Delta S_m=$~10~J/kg.K at 5 Tesla field change \cite{DankovPRB98}. However, the high cost of Gd makes it undesirable for the application of commercial magnetic refrigerator. As the Heusler alloys also have the potential for magnetic refrigerator applications, extensive efforts have been made on the study of magnetoclaoric properties in these materials \cite{PlanesJPCM09,OvichiJAP15,SokolovskiyPRB15}. In particular Ni--Mn--Z (Z = In, Sn, Sb) shape memory Heusler alloys, which show first order phase transition (FOPT), have been widely explored for the magnetic refrigerator applications \cite{TegusNat02, BassoPRB12, KrenkeNM05}. These alloys exhibit both conventional and inverse MCE i.e. negative and positive magnetic entropy change ($\Delta S_m$), respectively. This remarkable feature of these shape memory alloys is because of the coexistence of both structural and magnetic transitions. These properties are strongly correlated to each other and can easily be tuned with sample composition \cite{SahooJAP11, FujitaPRB01, WadaAPL01}. Although, the peak value of $\Delta S_m$ is found to be very high in FOPT alloys which is desirable for better cooling efficiency, but the presence of thermal and magnetic hysteresis limits their use in practical applications \cite{HalderJAP11, PecharskyPRL97}. 

In recent years, the magnetocaloric properties have been explored in Co-based Heusler alloys. These alloys show a second order phase transition (SOPT), and have advantages of broad operating temperature range \cite{Kuz07} and absence of thermal and/or magnetic hysteresis \cite{PandaJALCOM15, WangJAP09}. In order to better understand the magnetic transition in these alloys, the study of critical phenomena in the vicinity of the phase transition will have a great significance. The critical behavior across the paramgnetic to ferromagnetic (PM--FM) transition of perovskite manganites in double exchange model was first explained by long range mean-field theory. A few detailed studies of the critical phenomena in double exchange ferromagnetic La$_{1-x}$A$_x$MnO$_3$ ($A=$ Sr, Ca) have been reported in literature \cite{GhoshPRL98, MillisPRL95, MoriPRL98}. Moreover, the analysis of critical exponents for appropriate model proves very informative in examining the magnetic interactions and establishing a correlation between the critical exponents and exchange integral responsible for ferromagnetic transition. The critical exponents near the PM--FM transition in the austenite phase of Ni$_{43}$Mn$_{46}$Sn$_8$X$_3$ ($X=$ In and Cr) Heusler alloys have been analysed within the framework of long range mean-field theory \cite{NanJMMM18}. The recent results show deviation of critical exponents from the long range mean-field model, which is explained with two different behaviors that belonging to two different universality classes \cite{ZhangIEEE12, RajiJAP15}. Although, a close relationship between the critical behavior and magnetic exchange interaction has been perceivd, a fundamental question about the universality class for PM--FM phase transition in these materials is still under debate, and hence need further investigation.
 
Here we investigate the MCE in the vicinity of PM--FM transition temperature of Co$_2$Cr$_{1-x}$Mn$_x$Al alloys ($x=$ 0.25, 0.5, 0.75). The powder x-ray diffraction patterns indicate all the samples to be single phase, and the $x=$ 0.75 sample with L2$_1$ ordering. Our results mainly focus on the magnetic properties of these alloys and highlight the facts about the enhancement of magnetic moment, T$_{\rm C}$ and magnetic entropy change ($\Delta S_m$) with Mn substitution for multi-stage magnetic refrigeration applications. These samples show relatively high T$_{\rm C}$, which is very useful for technological applications and their structural and magnetic properties can be tuned easily/systematically. Therefore, understanding the critical behavior in Co$_2$Cr$_{1-x}$Mn$_x$Al would provide the information about the nature of the magnetic phase transition and spin interaction in these systems. We have performed a detailed analysis and extracted the critical exponents for the $x=$ 0.75 sample to understand the magnetocaloric properties. We found that the obtained parameters using modified Arrott plot (MAP) and Kouvel-Fisher (KF) method are very close to those expected for the mean field theory. The decay in the interaction distance suggests for the existence of the long-range FM interactions in these samples. Also, we have discussed the field dependent relative cooling power (RCP) in relation to the critical behavior analysis for the $x=$ 0.75 sample.

\section{\noindent ~Experimental Details}

The polycrystalline samples of Co$_2$Cr$_{1-x}$Mn$_x$Al ($x=$ 0.25, 0.5, 0.75) have been prepared by melting the constituent elements using the arc furnace, from Centorr vacuum industries, USA. The proper stoichiometric amount of starting materials of Co, Cr, Mn, Al (from Alfa Aesar or Sigma Aldrich) with purity of $\ge$ 99.9\% were melted on water cooled Cu hearth in the flow of inert argon. Further, we flipped the solid ingot and the melting process was repeated 4--5 times for a better homogeneity. After melting the weight loss was found to be less than 0.8\%. Then, the ingots were wrapped in the Mo sheet and sealed into the quartz tube under vacuum and annealed at 575~K for 10 days. The powder x-ray diffraction (XRD) measurements of Co$_2$Cr$_{1-x}$Mn$_x$Al ($x=$ 0.25, 0.5, 0.75) have been done with Cu K$\alpha$ ($\lambda$ = 1.5406~$\rm \AA$) radiation at room temperature and the XRD patterns were refined by Rietveld method using FullProf package, where the background was fitted using linear interpolation between the data points. The magnetic measurements were performed using a vibrating sample magnetometer (VSM) option of a physical property measurement system (PPMS, Quantum Design, Inc.) at the core lab for quantum materials at Helmholtz-Zentrum Berlin. 

\section{\noindent ~Results and Discussion}

\begin{figure}[h]
\includegraphics[width=3.2in]{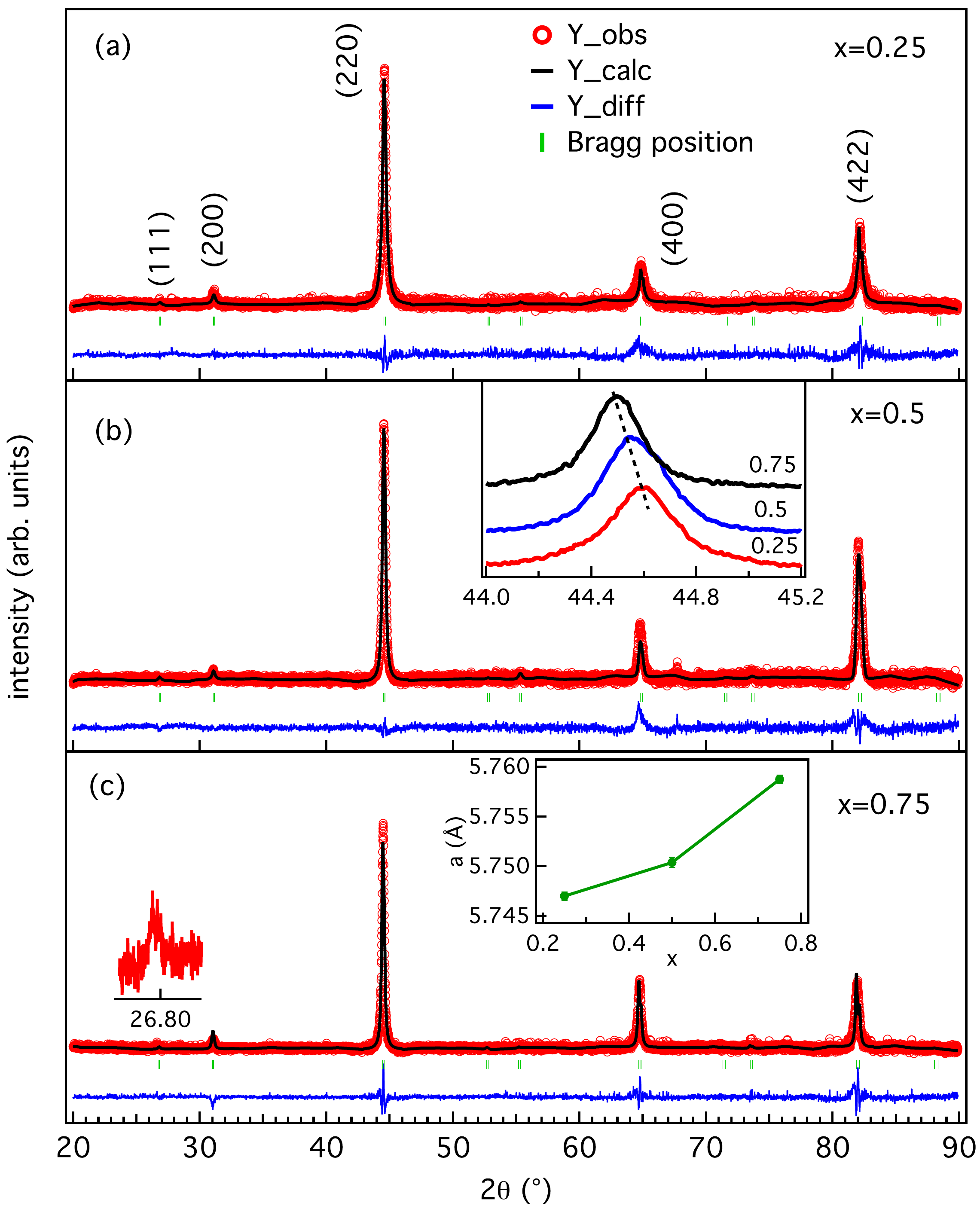}
\caption{The XRD patterns along with Rietveld refinement for the $x=$ 0.25 (a), 0.5 (b), and 0.75 (c) samples. Insets in (b) and (c) show shift in 220 Bragg peak position and variation of lattice constant $a$ with $x$, respectively, another inset in (c) is highlighting the presence of (111) Bragg peak at $\approx $ 26.7$\degree$.}
\label{fig1}
\end{figure}

The X$_2$YZ Heusler alloys crystallize in the cubic structure (L2$_1$) with space group $Fm\bar{3}m$. This unit cell consists of four interpenetrating bcc sublattices with X atoms positioned at (1/4, 1/4, 1/4) and (3/4, 3/4, 3/4), Y atom at (0, 0, 0) and Z atom at (1/2, 1/2, 1/2) sites \cite{GrafPSSC11}. We present the room temperature XRD patterns of Co$_2$Cr$_{1-x}$Mn$_x$Al in Figs.~\ref{fig1}(a--c) along with the Rietveld refinement using L2$_1$ structure and $Fm\bar{3}m$ (225) space group. We notice that (111) Bragg peak is not clearly visible in $x=$ 0.25 and 0.5 samples. On the other hand, for the $x=$ 0.75 sample we observed (111) Bragg peak at 2$\rm \theta \approx $ 26.7$\degree$, as shown in the inset of Fig.~\ref{fig1}(c). In general, the presence of (111) and (200) Bragg peaks in XRD confirms the L2$_1$ type ordering in full Heusler alloys i.e., all the atoms are occupying their own positions. 
However, absence of only (111) Bragg peak indicates the B2 type atomic ordering in the system, which generally appears due to random occupation between Cr/Mn and Al atoms. The intensity ratio between (111) and (200) peaks can give idea about the degree of disorder present in the system. The refinement and measured XRD patterns show that Mn substitution at Cr site increases the degree of ordering \cite{AlhajJAP12} and confirm the L2$_1$ structure in the $x=$ 0.75 sample. Moreover, we observed the shift in the (220) Bragg peak towards the lower 2$\theta$ value with increasing the $x$, as shown in the inset of Fig.~\ref{fig1}(b) and consequently lattice constant $a$ increases slightly with the Mn substitution (the $x$ value), see the inset of Fig.~\ref{fig1}(c). This increase in $a$ is contrary to the expectation as the radius of Mn is smaller than Cr.

\begin{figure}
\includegraphics[width=3.5in]{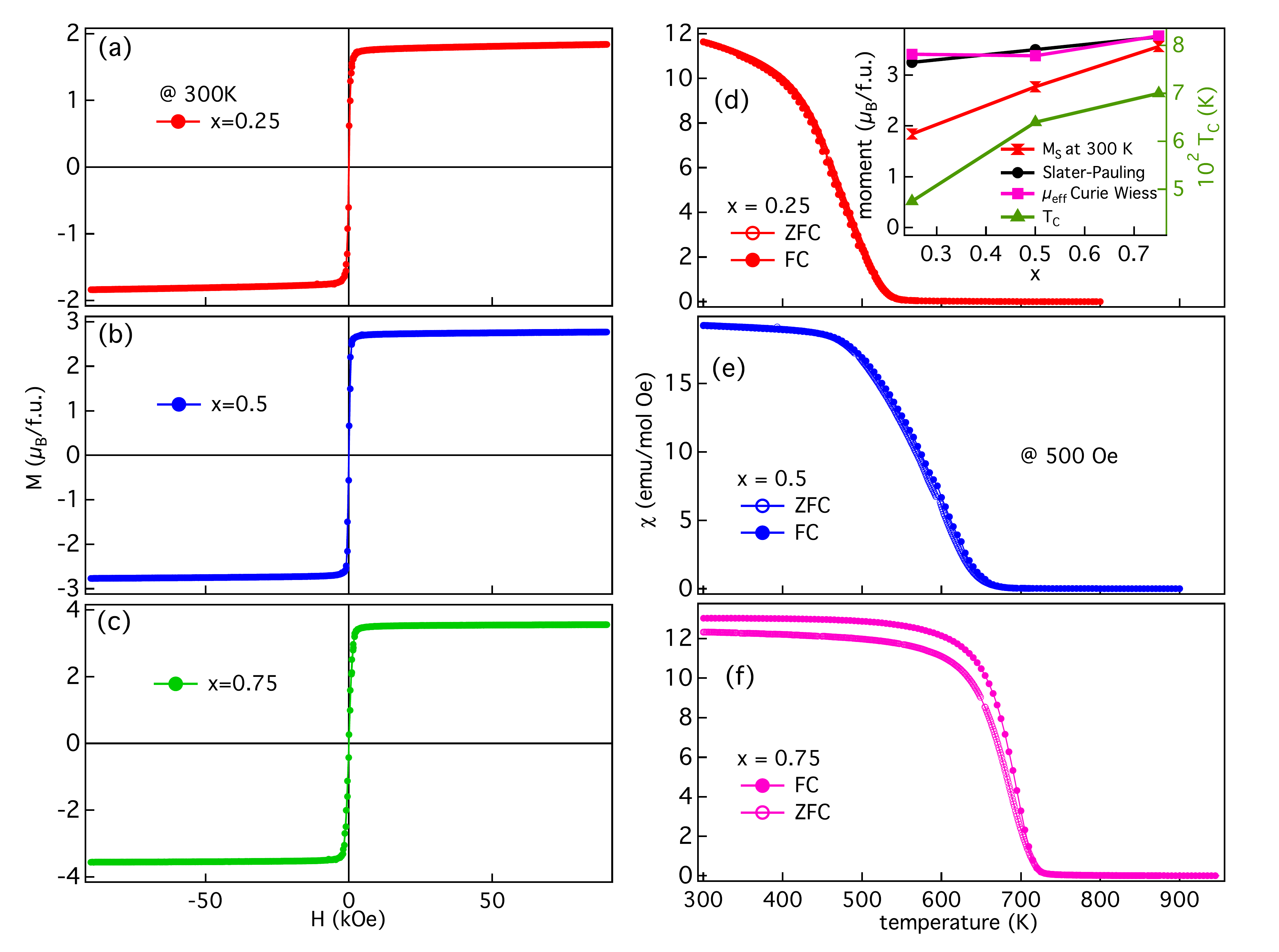}
\caption{(a--c) Isothermal magnetization M as a function of field H measured at 300~K and (d--f) Zero-field-cooled (ZFC) and field-cooled (FC) magnetic susceptibility $\chi$ with temperature measured at 500~Oe field for the Co$_2$Cr$_{1-x}$Mn$_x$Al ($x=$ 0.25, 0.5, 0.75) samples. Inset in (d) shows the variation of magnetic moment calculated from different methods (SP rule, saturation magnetization, Curie Weiss law) and T$_{\rm C}$ with $x$.}
\label{fig2}
\end{figure}

Figures~\ref{fig2}(a--c) show the isothermal magnetization M(H) data of Co$_2$Cr$_{1-x}$Mn$_x$Al for the $x=$ 0.25, 0.5 and 0.75 samples measured at 300~K. Interestingly, these samples exhibit a soft ferromagnetic behavior and the total magnetic moment calculated from isothermal magnetization at 300~K increases with the Mn substitution, as expected from the Slater-Pauling (SP) equation \cite{GalanakisPRB02, AlhajJAP12} $M_t=$ (N$_V-$24), where $M_t$ and $N_V$ are magnetic moment in Bohr magneton per formula unit ($\mu_{\rm B}/f.u.$) and the number of valence electrons per formula unit, respectively. An increase in Mn concentration would imply an increased number of valence electrons in the sample (as Mn has one more valence electron than Cr); hence, the calculated magnetic moment according to the SP equation is expected to increase, as shown in the inset of Fig.~2(d). For comparison, we have plotted the magnetic moment values (saturation magnetization, M$_{\rm S}$) from the isothermal magnetic measurements in FM state at 300~K. We note that the experimental values show an increasing behavior as expected from the SP rule. However, the lower values as compare to the SP rule for the $x=$ 0.25 and 0.5 samples might be due to the presence of atomic disorder, which is suggested to reduce the ferromagnetic ordering (moment) in these Heusler alloys \cite{KudryavtsevPRB08, MiuraPRB04}. On the other hand, for the $x=$ 0.75 sample, the M$_{\rm S}$ is almost same as calculated from the SP equation. This is also consistent with the fact that we observed (111) Bragg peak, which confirms L2$_1$ ordered state in the $x=$ 0.75 sample. The thermomagnetic ZFC (zero field cooled) and FC (field cooled) curves have been shown in Figs.~\ref{fig2}(d--f) for all the samples. The consistent decrease in magnetization with temperature and absence of thermal hystersis indicate a second order PM to FM phase transition. The reported T$_{\rm C}$ of the parent sample Co$_2$CrAl is close to the room temperature (315~K) \cite{HusmannPRB06, KourovPSS13}. It should be noted that the Mn substitution at the Cr site in Co$_2$CrAl strongly influences the Curie temperature (T$_{\rm C}$) and $\mu_{eff}$ obtained from the Curie-Weiss law.  As shown in the inset of Fig.~\ref{fig2}(d), the T$_{\rm C}$ increases with the Mn concentration. This indicates that the substitution of Mn at Cr site increases the T$_{\rm C}$ from 315~K ($x=$ 0 sample) to about 720~K for the $x=$ 0.75 sample. Interestingly, the absence of thermal hysteresis in M--T data [Figs.~\ref{fig2}(d--f)] is the signature of SOPT in these samples (detailed analysis discussed later), which makes them a favorable candidate and motivate to investigate the magnetocaloric properties. The influence of Mn substitution on the T$_{\rm C}$ is beneficial for controlling the working temperatures for spintronic and magnetic refrigerant applications.

\begin{figure}
\includegraphics[width=3.4in]{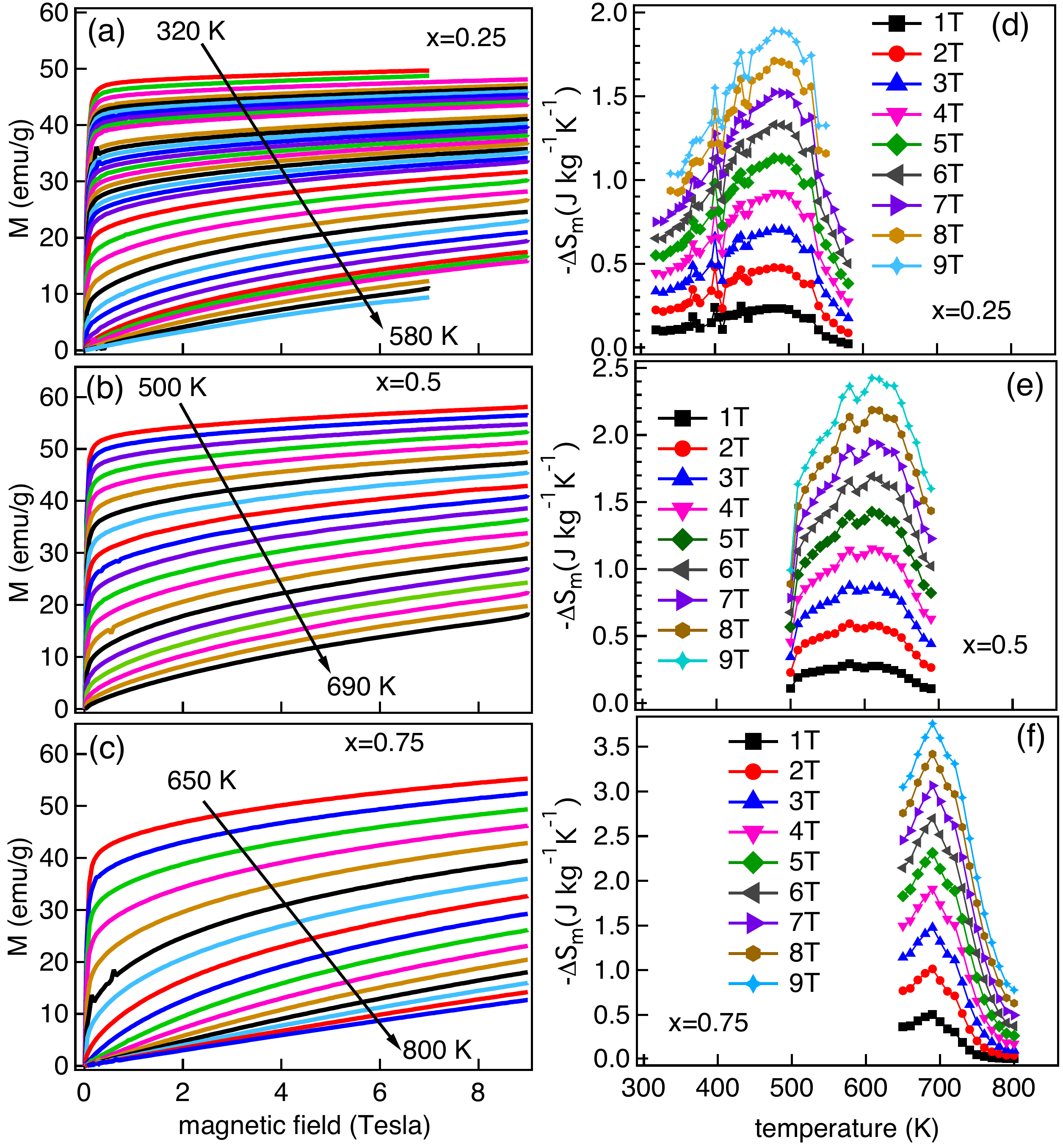}
\caption{Isothermal magnetization M(H), change in magnetic entropy ($\Delta$S$_m$) with temperature for the Co$_2$Cr$_{1-x}$Mn$_x$Al (a, d) $x=$ 0.25, (b, e) $x=$ 0.5 and (c, f) $x=$ 0.75 samples.}
\label{fig3}
\end{figure} 

The isothermal magnetization data measured across the transition temperature (T$_{\rm C}$) and presented in Figs.~3(a--c) are used to investigate the magnetocaloric effect (MCE) of Co$_2$Cr$_{1-x}$Mn$_x$Al ($x=$ 0.25, 0.5, 0.75). The change in magnetic entropy ($\Delta S_m$) with temperature has been obtained at different magnetic fields from 1 to 9~Tesla using the following Maxwell's relation:
\begin{equation}
\Delta S_m= \int_{0}^{H} \bigg(\frac{\delta M(H,T)}{\delta T}\bigg)_H dH
\label{eq1}
\end{equation}
Figs.~\ref{fig3}(d--f) show the variation of entropy change ($\Delta S_m$) with temperature across T$_{\rm C}$ upto 9~Tesla field strength for Co$_2$Cr$_{1-x}$Mn$_x$Al ($x=$ 0.25, 0.5, 0.75) samples. All the samples show a peak in $\Delta S_m$ around the transition temperature. We can clearly see the increase in maximum ($\Delta S_m^{max}$) value with increase in the $x$ (Mn concentration). Although, the peak value of $\Delta S_m$ is lower in SOPT materials as compared to FOPT materials, no hysteresis loss is observed in SOPT materials and hence their refrigeration capacity (RC) can be more useful, discussed in detail later. The calculated values of $\Delta S_m^{max}$ are 1.9 and 2.4~J/kg.K at 9 Tesla for the $x=$ 0.25 and 0.5 samples, respectively. However, highest $\Delta S_m^{max}$ value was found for the $x=$ 0.75 sample, which is 3.8~J/kg.K at 9 Tesla. We note that the highest value of $\Delta S_m^{max}$ for the $x=$ 0.75 sample is relatively lower compare to the other known SOPT materials like expensive Gd \cite{DankovPRB98}. On the other hand, high value of $\Delta S_m^{max}$ can be found in transition metal based intermetallic alloys, for example Ni$_2$MnIn (6.3~J/kg.K at 5 Tesla) \cite{SinghAM16}, Pt-doped Ni-Mn-Ga (7~J/kg.K at 5 Tesla) \cite{SinghAPL14},  MnFeP$_{0.45}$As$_{0.55}$ (18~J/kg.K at 5 Tesla) \cite{TegusNat02}, etc. However, the increasing behavior of $\Delta S_m^{max}$ and broad temperature range is suitable for Ericsson-cycle magnetic refrigerant applications \cite{Takeya94}. In this direction, it is important to note that understanding the magnetic properties and MCE makes Co$_2$-based Heusler alloys useful for both spintronics and multi-stage magnetic refrigerators.

\begin{figure}[h]
\includegraphics[width=3.6in]{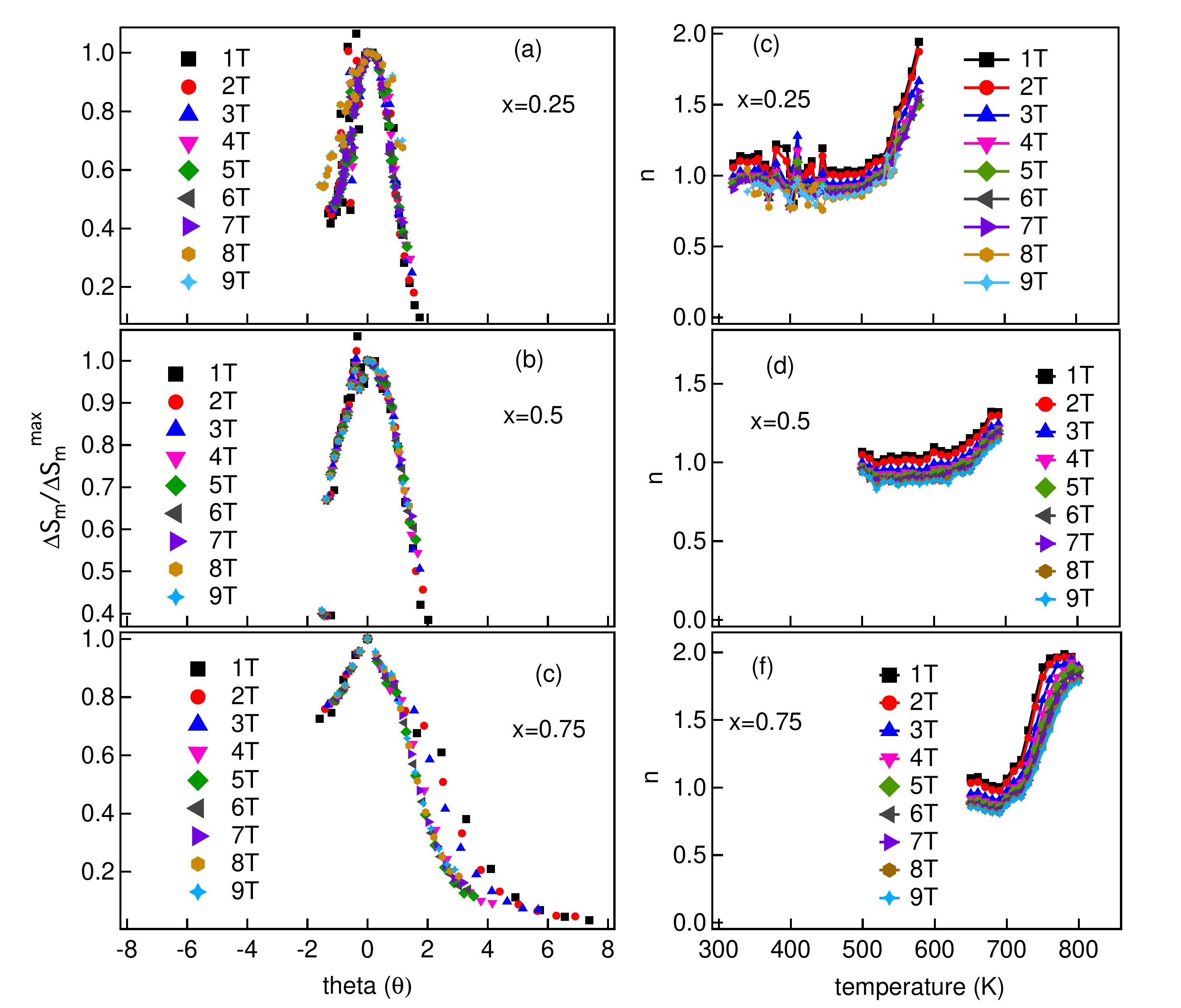}
\caption{(a--c) The scaled change in entropy with scaled temperature and (d--f) the variation of $n$ with temperature obtained at different applied magnetic fields for the Co$_2$Cr$_{1-x}$Mn$_x$Al ($x=$ 0.25, 0.5, 0.75) samples.}
\label{fig4}
\end{figure} 

Franco and Conde first proposed a method to check the nature of magnetic phase transition by the MCE. According to this method, for the second order magnetic phase transition (SOPT), all the entropy curves at different applied fields should collapse into a single curve after scaling  \cite{MarcelaPRB2010}. In these curves the entropy has been normalized and  temperature axis is scaled into $\theta$, which is defined as \cite{FrancoJPCM2008}:
\begin{equation}
\begin{aligned}
\theta = 
\begin{cases}
-(T-T_{pk})/ (T_{r1}-T_{pk}); 	\quad T\leq{T_c}\\
(T-T_{pk})/ (T_{r2}-T_{pk}); 	\quad T>{T_c}
\end{cases}
\end{aligned}
\label{eq2}
\end{equation}
where $T_{r1}$ and $T_{r2}$ are the reference temperatures below and above the T$_C$, respectively, corresponding to certain $f$ value and T$_{pk}$ is the temperature corresponding to peak value of $\Delta S_m$. In our case $f=\Delta S_m/\Delta S_m^{max}=$ 0.8 has been chosen for all the samples. Figures.~\ref{fig4}(a--c) show the variation of $\Delta S_m/\Delta S_m^{max}$ as a function of scaled temperature $\theta$. As evident from Figs.~\ref{fig4}(a--c), all the curves at different applied fields upto 9~Tesla collapses into a single curve, which is the signature of SOPT in Co$_2$Cr$_{1-x}$Mn$_x$Al ($x=$ 0.25, 0.5, 0.75) samples.

Further analysis of the MCE in Co$_2$Cr$_{1-x}$Mn$_x$Al samples, in the case of mean field model, gives a direction to check the dependency of $\Delta S_m$ with applied magnetic field. The $\Delta S_m$ has a power law dependency with the field as $|\Delta S_m|= a (H)^n$, where $a$ is a constant and $n$ is an exponent of magnetic entropy change that depends on the magnetic state of the material. From the entropy change curves, $n$ at different applied fields across the transition temperature can be calculated as \cite{FrancoJMMM09}:
\begin{equation}
n= \frac{d(ln |\Delta S_m|)}{d(lnH)}
\label{eq0}
\end{equation}
The dependency of $n$ on temperature have been thoroughly explored for SOPT materials in earlier reports \cite{FrancoJPCM2008}. For a small field, the value of $n$ can not be considered due to multi domain state in the sample. The temperature and field dependent $n$ is shown in Figs.~\ref{fig4}(d--f), where at temperatures well below and above the $T_{\rm C}$, the value of $n$ found to be equal to 1 and 2, respectively. This means in this temperature range across the transition $n$ changes from 1 to 2 i.e., $n=$ 1 at fairly ferromagnetic region, attains its minimum value at T$_{\rm C}$ and $n=$ 2 in the paramagnetic region as a consequence of Curie-Weiss law, unlike FOPT type materials where $n>$ 2 \cite{LawNatureCom18}. This analysis confirms the nature of second order type magnetic transition in the present study and is also consistent with a recent interesting study by Law {\it et al.} \cite{LawNatureCom18}.

\begin{figure}
\includegraphics[width=3.5in]{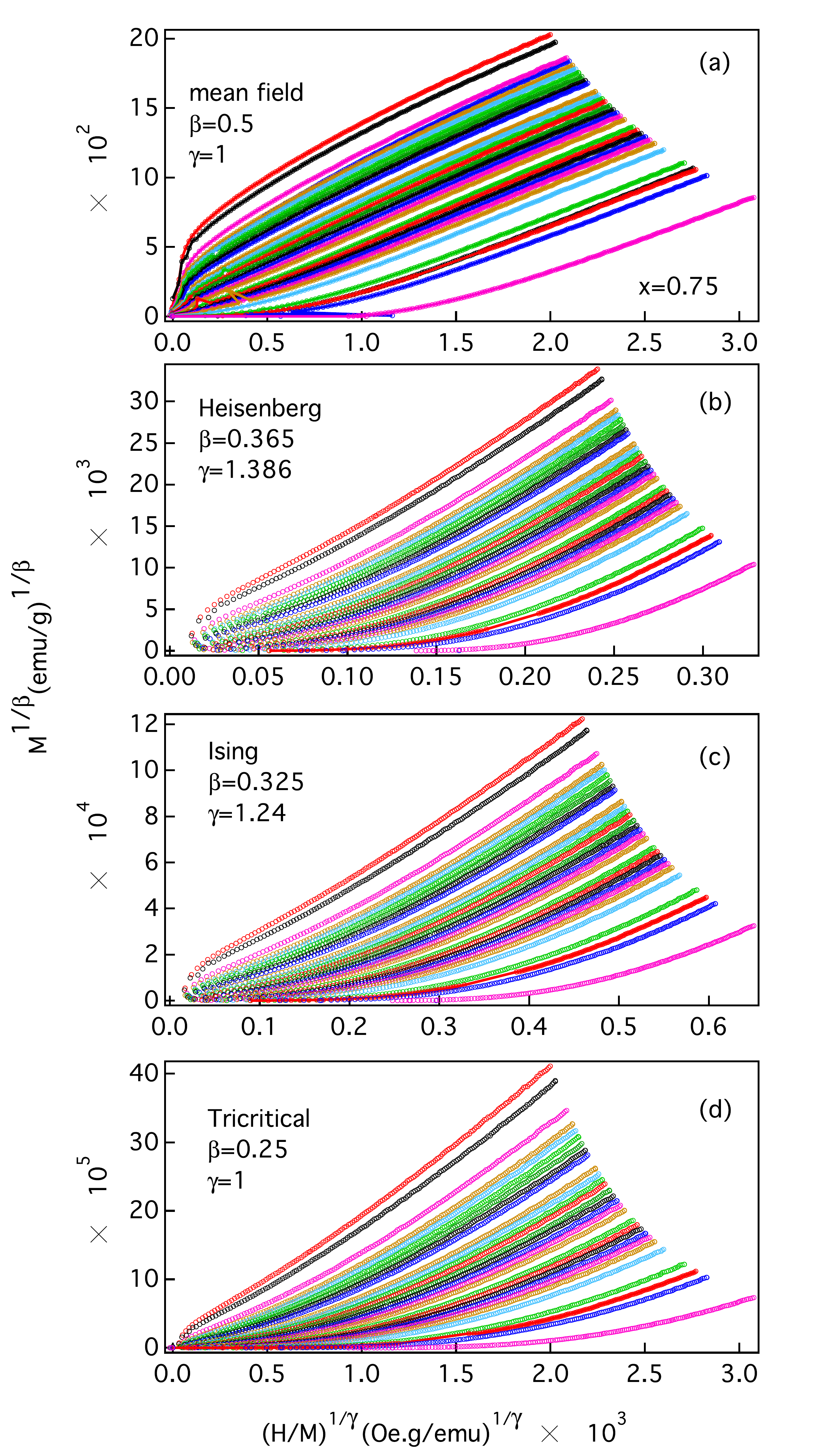}
\caption{Isotherms $M^{1/\beta}$ vs $(H/M)^{1/\gamma}$ using (a) mean field, (b) Heisenberg, (c) Ising and (d) Tricritical mean field models for the $x=$ 0.75 sample.} 
\label{fig5}
\end{figure} 

\begin{figure}
\includegraphics[width=3.5in]{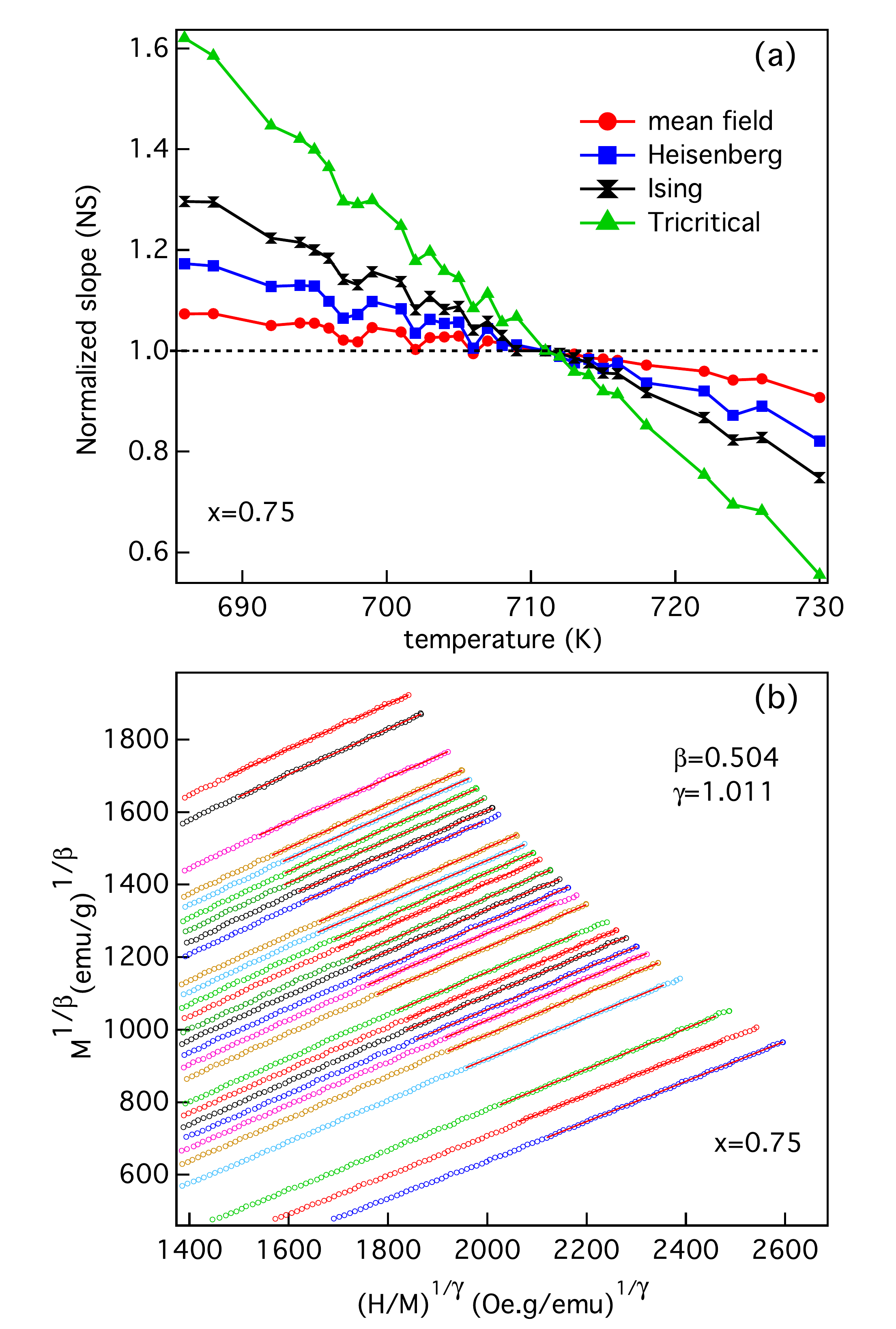}
\caption{(a) The temperature dependence of normalized slope (NS) of four different models and (b) the modified Arrott plot ($M^{1/\beta}$ vs $(H/M)^{1/\gamma}$) of isotherms for the $x=$ 0.75 sample.}
\label{fig6}
\end{figure}

The magnetization study shows the magnetic transition from PM to FM state, as also shown by the M as a function of H for selected temperatures near the phase transition, T$_{\rm C}$. In general, we get more information about the PM--FM transition from Arrott plots, where M$^2$ vs H/M curves should have the parallel lines in high field region for the mean field model. The Arrott plot is the alternative method to get an idea about the nature of phase transition. According to Banerjee criterion positive and negative slope of straight line fit in Arrott plot is the signature of SOPT and FOPT nature, respectively \cite{BanerjeePL64}. From the Arrott plot and Banerjee criterion, we can conclude that Co$_2$Cr$_{1-x}$Mn$_x$Al ($x=$ 0.75) exhibits a second order phase transition. Moreover, the magnetization curves near PM--FM phase transition temperature should follow the Arrott-Noakes equation of state \cite{Arrott67}:
\begin{equation}
\bigg({\frac{H}{M}}\bigg)^{\frac{1}{\gamma}}= a \frac{(T-T_c)}{T}+bM^{\frac{1}{\beta}}
\label{eq24}
\end{equation}
where $a$, $b$ are constants. The $\beta$, $\gamma$ and T$_{\rm C}$ are critical exponents and Curie temperature, respectively. This equation gives a set of parallel straight lines with correct combination of $\beta$, $\gamma$ and T$_{\rm C}$.

In order to understand the nature of magnetic phase transition near T$_{\rm C}$, the nature of the ordering and the MC behavior of Co$_2$Cr$_{1-x}$Mn$_x$Al, we have performed detailed analysis to extract the values of critical exponents ($\beta$, $\gamma$ and $\delta$) for the $x=$ 0.75 sample. Here, the critical exponents $\beta$, $\gamma$ and $\delta$ are correspond to the saturation magnetization M$_S$(T), the inverse initial magnetic susceptibility $\chi_0^{-1}$ and the critical magnetization at T$_{\rm C}$, respectively \cite{Arrott67}. The critical exponents $\beta$ and $\gamma$ can be calculated using the following equations \cite{JamesPRB1964, FisherRPP67}:
\begin{equation}
M_S(T) = M_0(-\epsilon)^\beta; \quad \epsilon <0, T<T_c
\label{eq4}
\end{equation}
\begin{equation}
\chi_0^{-1} = \frac{h_0}{M_0}(\epsilon)^\gamma; \quad \epsilon >0, T>T_c
\label{eq5}
\end{equation}
where $\epsilon$= [(T-T$_{\rm C}$)/T$_{\rm C}$] and $h_0/M_0$ are the reduced temperature and the critical amplitude, respectively. More details about the physical meaning of these critical amplitudes can be found in refs.~\cite{FisherRPP67, SiruguriJPCM96}.

As there can be different types of magnetic interactions corresponding to the spin-dimensionality $d$ (discussed later) \cite{FisherPRL72, FisherRMP74}, the Heisenberg model shows isotropic magnetic coupling, when $d=$ 3, whereas $d=$ 1 belongs to the Ising model and corresponds to an anisotropic magnetic interaction \cite{FisherRMP74}. The tricritical model is applicable to the multiple phase system \cite{Huang87}. These four models have been considered for initial assumption of $\beta$ and $\gamma$ values i.e. mean field model ($\beta=0.5$, $\gamma= 1$), Heisenberg model ($\beta=0.365$, $\gamma= 1.386$), Ising model ($\beta=0.325$, $\gamma= 1.24$) and tricritical model ($\beta=0.25$, $\gamma= 1$). First we use these values of $\beta$ and $\gamma$ to construct modified Arrott plots [M$^{1/\beta}$ vs (H/M)$^{1/\gamma}$], which for all the four models are shown in Figs.~\ref{fig5}(a--d) for the $x=$ 0.75 sample. The best model can be decided by two criteria: (1) all the lines in the high field region in M$^{1/\beta}$ vs (H/M)$^{1/\gamma}$ plot should be straight and (2) they should be parallel to each other. The plots of  M$^{1/\beta}$ vs (H/M)$^{1/\gamma}$ should exhibit a set of straight lines parallel to each other with same slope S(T) around T$_{\rm C}$. The line which passes through origin will define the Curie temperature (T$_{\rm C}$). Although, all the models have straight lines in high field region, the slope is not same near T$_{\rm C}$. To further check the second criteria normalized slope ($N S$) has been calculated. The $N S$, expressed as $N S=$ S(T)/S(T$_{\rm C}$), has been calculated at each temperature near T$_{\rm C}$, where $S(T)$ is the slope of M$^{1/\beta}$ vs (H/M)$^{1/\gamma}$ at temperature T and S(T$_{\rm C}$) is the slope at T$_{\rm C}$. For a best fit model, the value of $N S$ should be almost equal to one.

\begin{figure*}
\includegraphics[width=7.3in]{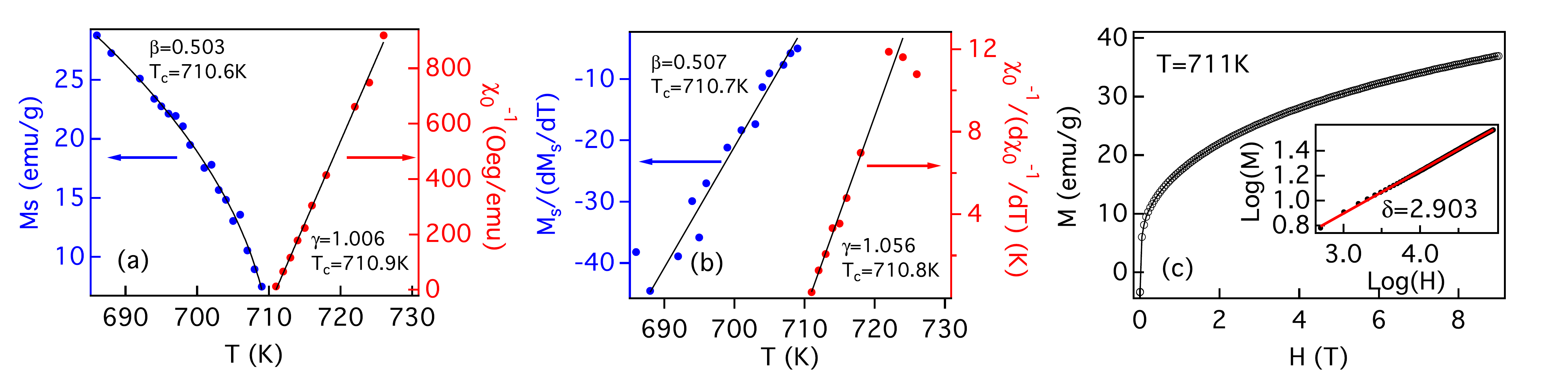}
\caption{(a) The plot of M$_{\rm S}$ (left axis) and $\chi_{0}^{-1}$ (right axis) as a function of temperature, (b) Kouvel-Fisher plot of M$_{\rm S}$ (left axis) and $\chi_{0}^{-1}$ (right axis) as a function of temperature, and (c) M vs H plot at 711~K for the $x=$ 0.75 sample. The inset in (c) is the same plot in log-log scale with the linear fit (solid red line).}
\label{fig7}
\end{figure*}

Figure~\ref{fig6}(a) shows the temperature variation of $N S$ for the $x=$ 0.75 sample fitted for modified Arrott plots. We find that the critical behavior of the Co$_2$Cr$_{0.25}$Mn$_{0.75}$Al sample is described best by the mean field model. Note that initially we constructed a modified Arrott plot using equation \ref{eq24} with a set of $\beta$ and $\gamma$ values according to the mean field model. Then the best values of $\beta$ and $\gamma$ were obtained by iteration method \cite{Zhang12} using equations \ref{eq4} and \ref{eq5}. The M$_S$ (T, 0) and $\chi_0^{-1}$(T, 0) for zero field are estimated from the linear extrapolation of M$^{1/\beta}$ vs H/M$^{1/\gamma}$ lines in high field region to the positive intercept on M$^{1/\beta}$ (below T$_c)$ and H/M$^{1/\gamma}$ (above T$_{\rm C}$) axes, respectively and the fitting of the curves with equations \ref{eq4} and \ref{eq5} yields new $\beta$, $\gamma$ and T$_{\rm C}$. This process is iterated until we get the stable values of $\beta$ and $\gamma$. After few iterations the modified Arrott plot with so obtained $\beta=$ 0.504, $\gamma=$ 1.011 for the $x=$ 0.75 sample is shown in Fig.~\ref{fig6}(b), which forms the set of parallel lines in high field region. However, the lines in the low field region are curved due to different domains magnetized in different directions. To further check the critical exponents, the intercept of M$^{1/\beta}$ and (H/M)$^{1/\gamma}$ from Fig.~\ref{fig6}(b) gives M$_S$ and $\chi_0^{-1}$, respectively and fitting of M$_S$ vs T and $\chi_0^{-1}$ using equations \ref{eq4} and \ref{eq5} gives the new set of critical exponents i.e. $\beta=$ 0.502, $\gamma=$ 1.006, as plotted in Fig.~\ref{fig7}(a). The values of these critical exponents are quite close to the values obtained in Fig.~\ref{fig6}(b).

\begin{table*}[ht]
\caption{The critical exponents ($\alpha$, $\beta$, $\gamma$ and $\delta$) determined from modified Arrott plot, Kouvel-Fisher method and critical isotherm for Co$_2$Cr$_{0.25}$Mn$_{0.75}$Al. The values of these parameters for Co$_2$CrAl (from \cite{PandaJALCOM15}) and calculated from theoretically predicted universality classes have also been included for direct comparison.}
\vskip 0.1 cm
\centering
\begin{tabular}{c c c c c c c} 
\hline\hline 
Sample/Model & Reference(s) & Method(s) & $\alpha$ & $\beta$ & $\gamma$ & $\delta$\\ [0.5ex] 
\hline
Co$_2$CrAl 				  & \cite{PandaJALCOM15} & Modified Arrott plot &  & 0.488$\pm$0.003 & 1.144$\pm$0.004 & 3.336$\pm$0.005\\
 		    				  &  & Kouvel-Fisher method &  & 0.482$\pm$0.013 & 1.148$\pm$0.016 & 3.382$\pm$0.020\\
 		    				  &  & Critical isotherm &  &  &  & 3.401$\pm$0.004\\
Co$_2$Cr$_{0.25}$Mn$_{0.75}$Al    & This work & Modified Arrott plot &  & 0.503$\pm$0.025 & 1.006$\pm$0.050 & 3.004$\pm$0.140\\
 		   				 &  & Kouvel-Fisher method &  & 0.507$\pm$0.026 & 1.056$\pm$0.055 & 3.083$\pm$0.152\\
 		    				 &  & Critical isotherm &  &  &  & 2.903$\pm$0.003\\
Mean field			 & \cite{KaulJMMM85,Stanley} & Theoretical & 0 & 0.5 & 1.0 & 3.0\\
3D Heisenberg		 & \cite{KaulJMMM85,GuillouPRB80, FisherRMP74} & Theoretical & -0.115 & 0.365 & 1.386 & 4.80\\
3D Ising			 & \cite{KaulJMMM85,GuillouPRB80, FisherRMP74} & Theoretical & 0.11 & 0.325 & 1.241 & 4.82\\
Tricritical mean field			 & \cite{KimPRL02} & Theoretical &  & 0.25 & 1.0 & 5.0\\
\hline\hline
\end{tabular}
\label{table3} 
\end{table*}

Furthermore, we use Kouvel-Fisher method in order to validate the accuracy of these critical exponents determined from the modified Arrott plot, where the modified equations \ref{eq4} to \ref{eq5} are written below as per the Kouvel-Fisher method \cite{JamesPRB1964, FisherRPP67}:
\begin{equation}
\frac{M_S(T)}{dM_S(T)/dT} = \frac{T-T_c}{\beta}
\label{eq7}
\end{equation}
\begin{equation}
\frac{\chi_0^{-1}(T)}{d\chi_0^{-1}(T)/dT} = \frac{T-T_c}{\gamma}
\label{eq8}
\end{equation}
\begin{equation}
ln M(H,T_c) = \frac{ln H}{\delta}
\label{eq9}
\end{equation} 
Here the critical exponents ($\beta$ and $\gamma$) are obtained from the plots of $M_S$ (d$M_S/dT)^{-1}$ vs T and $\chi_0^{-1}$ (d$\chi_0^{-1}/dT)^{-1}$ vs T, as shown in Fig.~\ref{fig7}(b). A linear fit to these two curves yields critical exponents where the values of $1/\beta$ and $1/\gamma$ are determined from the respective slope according to the equations \ref{eq7} and \ref{eq8}, respectively, as shown in Fig.~\ref{fig7}(b). Also, an intercept on $x-$axis for each curve corresponds to the T$_{\rm C}$. In addition, according to the equation \ref{eq9}, the plot between ln(M) and ln(H) at T$_{\rm C}$ should be a straight line with slope of 1/$\delta$. In the inset of Fig.~\ref{fig7}(c), we show the straight line fit, which gives rise the value of $\delta=$ 2.903. Moreover, we have utilized another method to check the reliability of $\delta$ value using Widom scaling relation $\gamma - \beta (\delta -1)=$ 0. The obtained critical exponents from Kouvel-Fisher method are very close to the values determined from Arrott-Noakes equation of state, as listed in Table~I. The theoretically calculated parameters are also  listed in Table~I. 

\begin{figure}
\includegraphics[width=3.5in]{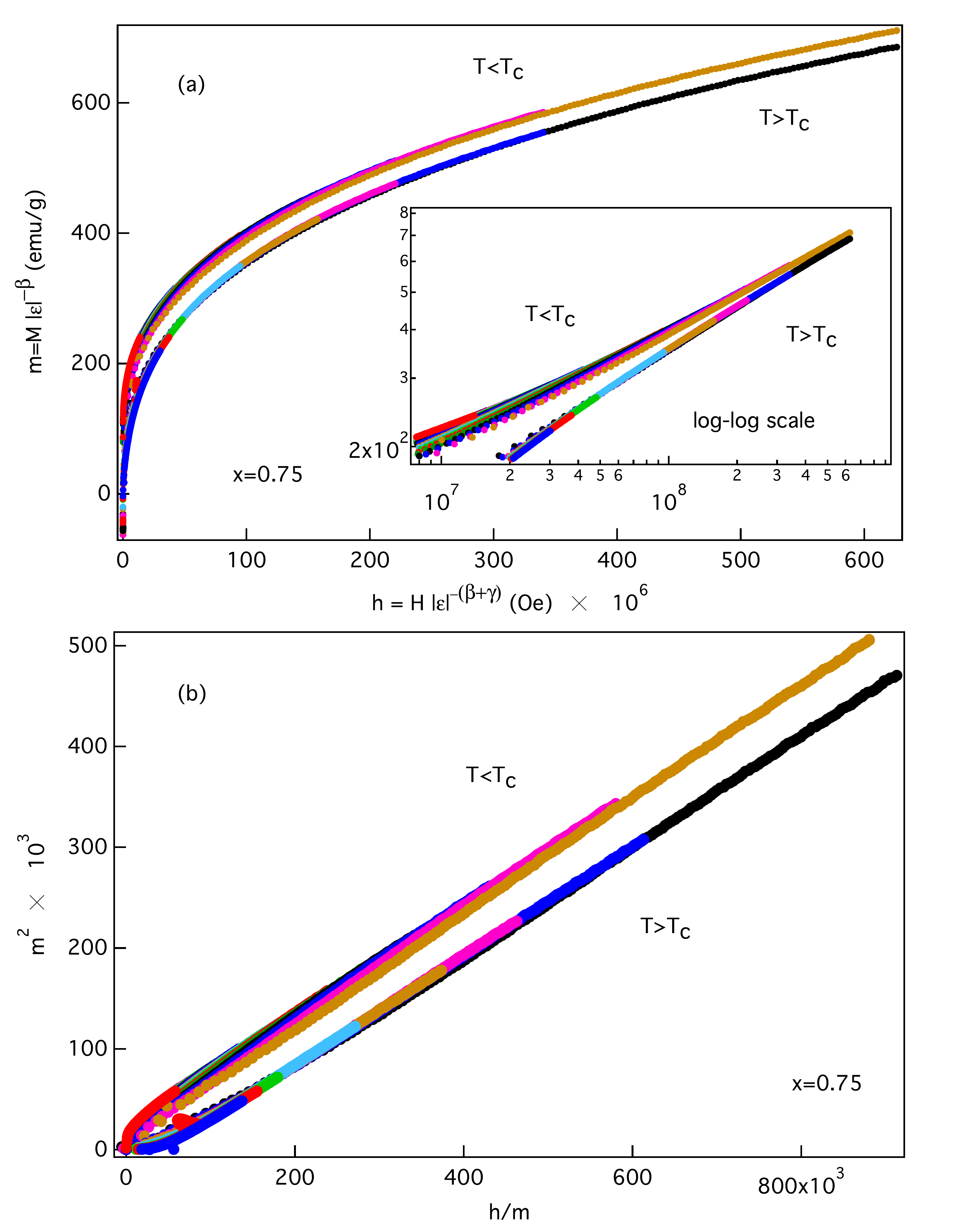}
\caption{(a) The renormalized magnetization ($m$) plotted (in left) as a function of renormalized field ($h$) and (b) the plot in form of $m^2$ vs $h/m$ for the $x=$ 0.75 sample. The different symbols correspond to different temperatures and the inset in (a) shows the log--log plot.}
\label{fig8}
\end{figure}

In the present case, the values of calculated critical exponents $\beta$ and $\gamma$ are close to the mean field model, which suggests for the presence of long range magnetic interactions in the system \cite{KaulJMMM85}. However, it is vital to further confirm this inference. Therefore, it is important to check the scaling equation of state with these critical exponents \cite{PramanikPRB09, HalderJAP11,YanSSC18}. The scaling equation of state which describes the relationship between $M$($H$,$\epsilon$), $H$ and T$_{\rm C}$ can be expressed as:
\begin{equation}
M(H,\epsilon) = \epsilon^{\beta}f_{\pm} (\frac{H}{\epsilon^{\beta+\gamma}})
\label{eq3}
\end{equation}
where $f_-$ and $f_+$ are regular analytical functions below and above T$_{\rm C}$, respectively. The equation~\ref{eq3} states that for the proper combination of $\beta$, $\gamma$ and T$_{\rm C}$, the plot of scaled magnetization $m=|\epsilon|^{-\beta} M(H, \epsilon$) as a function of scaled field $h=|\epsilon|^{-(\beta+\gamma)}H$ would generate the two distinct branches across the Curie temperature T$_{\rm C}$. Using equation~\ref{eq3}, the scaled $m$ vs scaled $h$ has been plotted in Fig.~\ref{fig8}(a) with the above calculated critical exponents ($\beta$, $\gamma$) and $T_{\rm C}$. It demonstrates that all the data have separated in two branches below and above $T_{\rm C}$, where different symbols correspond to different temperatures. The inset in Fig.~\ref{fig8}(a) is the same plot, but shown on a log-log scale. The reliability of the exponents and $T_{\rm C}$ has been further ensured with more rigorous analysis method by plotting $m^2$ vs $h/m$ [Fig.~\ref{fig8}(b)], where this plot falls in two different branches across the T$_{\rm C}$.

In the previous section we mentioned that the magnetic interactions in the studied samples are of extended type. Therefore, understanding the nature and range of spin interactions become an important part of this study. The extended type spin interaction is expressed in terms of exchange distance by $J(r)$, in which the long range exchange interaction decays as $J(r)\sim r^{-(D+\sigma)}$ and the short range exchange interaction decays as $J(r)\sim e^{-r/b}$, where $r$ is the distance, $D$ is the lattice dimensionality, $\sigma$ is the range of interaction and $b$ is special scaling factor \cite{FischerPRB02}. The $\sigma$ can be calculated by renormalization group approach using the following equation \cite{FisherRMP74, FisherPRL72}:
\begin{equation}
\begin{split}
\gamma = 1+ \frac{4}{D}\frac{d+2}{d+8} \Delta \sigma + \frac{8(d+2)(d-4)}{D^2(d+8)^2}\\
\times\bigg[1+\frac{2G(\frac{D}{2})(7d+20)} {(d-4)(d+8)}\bigg]\Delta \sigma^2
\label{eq12}
\end{split}
\end{equation}
where $d$ is the spin dimensionality, $\Delta \sigma = \sigma-\frac{D}{2}$, $G(\frac{D}{2})=3-\frac{1}{4}(\frac{D}{2})^2$. According to this equation for the experimental value of $\gamma$ one can calculate the value of $\sigma$. Then the obtained $\sigma$ value can be used to calculate the other exponents using the expressions: $\upsilon=\gamma/\sigma$, $\alpha=2-\upsilon D$, $\beta=(2-\alpha-\gamma)/2$ and $\delta=1+\gamma/\beta$, where $\upsilon$ is the correlation length exponent. In the present study, for $d>1$ and $D=$ 3, the $\beta$ and $\delta$ are close to our previous experimental results calculated by K--F method only for $\sigma$ close to 3/2, which is less than 2. For three-dimensional material ($D=$ 3) mean field model ($\beta$=0.5, $\gamma$=1, $\delta$=3.0), the exchange interaction decays slower than power law of $J(r)\sim r^{-(3+\sigma)}$, which is only valid when $\sigma<2$. It is known that $\sigma<2$ corresponds to the long range spin interaction and $\sigma>2$ indicates the short range interaction. This indicates that the exchange interaction, which belongs to the mean field model decays slower than $r^{-4.5}$. This analysis reveals the presence of long range spin interactions in the $x=$ 0.75 sample.

\begin{figure}
\includegraphics[width=3.6in]{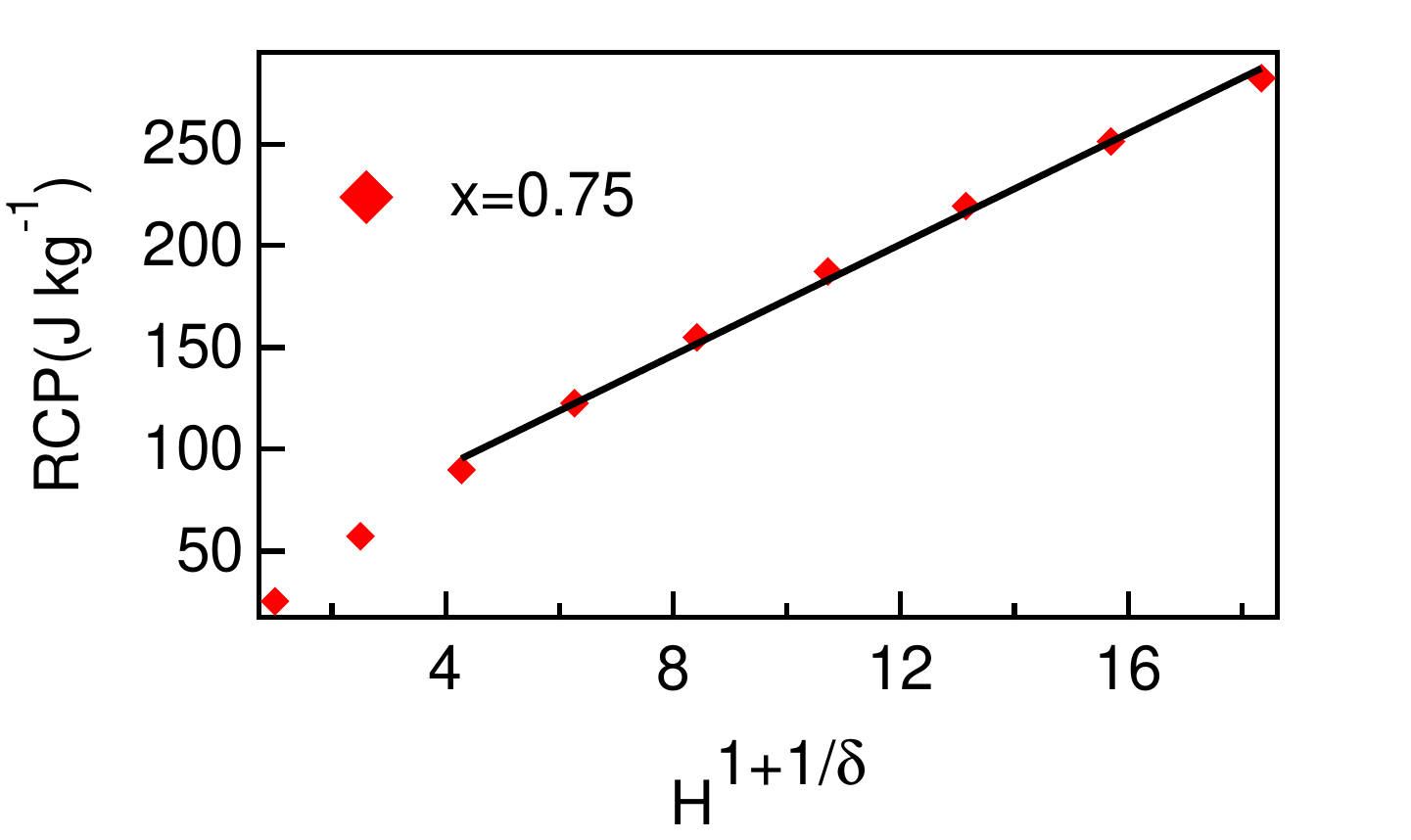}
\caption{The relative cooling power (RCP) with variation of $H^{1+\frac{1}{\delta}}$ for the $x=$ 0.75 sample, where the solid black line is a simple linear fit.}
\label{fig9}
\end{figure}

As we discussed before, from the variation of $n$ with temperature at various fields, we found the minimum value of $n \approx$ 0.80 (field independent) below T$_{\rm C}$, as shown in Figs.~\ref{fig4}(d--f), which is close to the mean field model \cite{OesterreicherJAP84}. The values of $n$ can also be obtained by the critical exponents $\beta$ and $\gamma$ at T$_{\rm C}$ i.e. by using $n$(T$_{\rm C}$) = 1 + ($\beta-$ 1)/($\beta+\gamma$). In fact, using this formula we estimate $n=$ 0.67, as predicted by Oesterreicher and Parker for the the mean field ($\beta=$ 0.5, $\gamma=$ 1, $\delta=$ 3) model \cite{OesterreicherJAP84}. In order to understand the relation between the critical behavior and the MCE, we now move to the analysis to check the refrigerator capacity, which depends on the temperatures of the cold and hot sinks of the cooling system and the change in isothermal entropy in the whole temperature range of the cooling cycle i.e., indicates the quantity of heat transferred between the cold and hot ends in the magnetic refrigerant in one ideal thermodynamic cycle \cite{PecharskyJAP01}. In this context, to find the suitability and efficiency of a magnetocaloric material, another parameter is defined as the relative cooling power (RCP), which is a property of the material, and its high value is desirable in cooling applications 
\cite{GschneidnerARM00, GschneidnerRPP05, PhanJMMM07, Buschow_book}. Here, we have investigated the RCP, which is the product of maximum entropy change and the temperature at full width half maximum (FWHM), as the equation below \cite{GschneidnerARM00, GschneidnerRPP05, PhanJMMM07, Buschow_book}:
\begin{equation}
RCP= |-\Delta S_m|\times \delta T_{FWHM}
\label{eq10}
\end{equation}
where $\delta T_{\rm FWHM}$ has been calculated from the magnetic entropy curves [see Fig.~3(f)] by fitting with the Gaussian function. The magnetic entropy peak is relatively broad in our case, which gives rise to a large temperature span ($\delta T_{\rm FWHM}$ $\approx$ 75~K at 9 Tesla). Notably, Engelbrecht {\it et al.} reported that a material with broad magnetic entropy peak is better than a material with a sharp peak for cooling applications, which suggests that a broad temperature distribution is more attractive for MCE \cite{EngelbrechtJAP10}. The value of RCP extracted from the above equation monotonically increases with field and attains a maximum value of about 282~J/kg at 9~Tesla (Fig.~\ref{fig9}), which is quite comparable with the other reported Co$_2$-based Heusler alloys \cite{PandaJALCOM15}. On the other hand, much higher value of RCP $\approx$400~J/kg at 5~Tesla has been reported for Gd \cite{MathewAPL11}. Here the challenge is to find a rare earth free material and therefore, NiMnX alloys have been extensively studied for better performing MCE \cite{HuAPL2000, ZavarehAPL15, LiSciRep15, ChenJALCOM18}. However, they mostly exhibit first order phase transition (FOPT), which causes a large thermal and magnetic hysteresis loss \cite{HuAPL2000, ZavarehAPL15, LiSciRep15, ChenJALCOM18}.
The present study makes the Co$_2$-based Heusler alloys a potential candidate for MCE even at a moderate RCP value where the transition temperature and maximum entropy change can be systematically controlled. 

In addition, to emphasize the relation between the critical exponents and the magnetocaloric effect, the RCP can be expressed with the magnetic field power law. The field dependence of power law with exponent $b$ is expressed as RCP$\propto$$H^b$ \cite{OumezzineJALCOM12}, where $b$ is defined with the critical exponent $\delta$ as $b={1+\frac{1}{\delta}}$. As listed in Table~I, from the calculations of the critical behavior analysis, we find the value of $\delta=$ 3.083 for Kouvel-Fisher method. Now taking into account this $\delta$ value, the plot of RCP vs field power ($H^{1+1/\delta}$) is shown in Fig.~\ref{fig9}. A straight line fit with this power law matches well with the experimental data points, which establishes a correlation between MCE and critical behavior for the $x=$ 0.75 sample. These experimental results and discussion suggest that these Heusler alloys exhibiting second order phase transition could be useful for high temperature cooling applications. Interesting to note that in a multi-stage magnetic refrigerator, the cooling from higher temperature is required in stage by stage manner \cite{KitanovskiIJR10}. These multi-stage refrigerators are very useful for industrial applications, where lower temperature in one stage acts as upper temperature for next stage \cite{KitanovskiIJR10, FrancoPMC18}. Therefore, these magnetic refrigerants with easily tunable magnetic moment and T$_{\rm C}$ can be useful at various operating temperatures. On the other hand, studies to find materials with considerable magnetocaloric effect near room temperature and investigation of the critical behavior with additional probes including heat capacity that can measure direct adiabatic temperature change would be highly desirable \cite{OesterreicherJAP84}.

\section{\noindent ~Conclusions}

In conclusion, we have performed a comprehensive analysis to study the MCE and the critical behavior of Co$_2$Cr$_{1-x}$Mn$_x$Al ($x=$ 0.25, 0.5, 0.75) Heusler alloys. The PM--FM phase transition is found to be a second order phase transition. The scaled change in magnetic entropy with variation of scaled temperature at different fields collapse into single curve, which also confirms the second order nature of the transition. The determined critical exponents have values corresponding to universality classes of mean-field theory, and match well with the values calculated from modified Arrott plot, Kouvel-Fisher method and critical isotherm. On the other hand, the scaling theory separates the field dependent magnetization in two branches, below and above the T$_{\rm C}$. The magnetic interactions in this alloy have been explained by the critical behavior and they follow the mean-field model with long range order magnetic interaction. The critical exponents estimated from the renormalization group approach are close to the obtained experimental values for $D=$ 3, $d>$ 1 that point towards the spin interaction of mean-field type with long range interaction and this interaction decays with distance $r$ as $J(r)\sim r^{-4.5}$. The field dependent relative cooling power (RCP) has been studied and discussed to relate with the critical behavior analysis for the $x=$ 0.75 sample. Interestingly, these high T$_{\rm C}$ materials can be used in multi-stage magnetic refrigerators where lower temperature in one stage acts as upper temperature for the next stage. Our study provides insight for making these Co$_2$ based Heusler alloys of practical use in both spintronic and magnetic cooling applications. However, the challenge remains to enhance magnetic entropy change ($\Delta S_m$) and the relative cooling power by maintaining second order phase transition near room temperature. 

\section{\noindent ~Acknowledgments}

Priyanka Nehla acknowledges the MHRD, India for fellowship. We thank the physics department, IIT Delhi for providing XRD facility and support. This work was financially supported by the BRNS through DAE Young Scientist Research Award to RSD with project sanction No. 34/20/12/2015/BRNS.

\end{document}